\documentclass[sigconf]{acmart}

\usepackage{booktabs}
\usepackage{multirow}
\usepackage{graphicx}
\usepackage{adjustbox}
\graphicspath{{figures/}}
\usepackage{tikz}
\usetikzlibrary{positioning,arrows.meta,shapes.geometric,fit,backgrounds,calc,shadows}
\usepackage{fontawesome5}

\AtBeginDocument{%
  }

\copyrightyear{2026}
\acmYear{2026}
\setcopyright{cc}
\setcctype{by-nc-nd}
\acmConference[SIGIR '26]{Proceedings of the 49th International ACM SIGIR Conference on Research and Development in Information Retrieval}{July 20--24, 2026}{Melbourne, VIC, Australia}
\acmBooktitle{Proceedings of the 49th International ACM SIGIR Conference on Research and Development in Information Retrieval (SIGIR '26), July 20--24, 2026, Melbourne, VIC, Australia}
\acmDOI{10.1145/3805712.3809970}
\acmISBN{979-8-4007-2599-9/2026/07}

\renewcommand\footnotetextcopyrightpermission[1]{} % 左下の著作権表示（DOI等）を一旦消す
\begin{document}

%%
%% Title
%%
\title{Same Image, Different Meanings: Toward Retrieval of Context-Dependent Meanings}

%%
%% Authors
%%
\author{Ayuto Tsutsumi}
\orcid{0009-0003-5380-7625}
\email{tsutsumi-ayuto@ed.tmu.ac.jp}
\affiliation{%
  \institution{Tokyo Metropolitan University}
  \city{Tokyo}
  \country{Japan}
}

\author{Ryosuke Kohita}
% \authornote{Corresponding author.}
\orcid{0009-0001-8414-9667}
\email{kohita_ryosuke@cyberagent.co.jp}
\affiliation{%
  \institution{CyberAgent}
  \city{Tokyo}
  \country{Japan}
}

\renewcommand{\shortauthors}{Tsutsumi et al.}

%%
%% Abstract
%%
\begin{abstract}
A scene of two people in the rain can convey hope and warmth in a reunion story or sorrow and finality in a farewell story.
We investigate this context-dependent nature of image meaning and its implications for retrieval.
Our key observation is that context dependency correlates with semantic abstraction: concrete elements (objects, actions) remain stable across contexts, while abstract elements (atmosphere, intent) shift with context.
We operationalize this as the L1--L4 framework, organizing image semantics from context-independent (L1) to maximally context-dependent (L4).
Using synthetic story contexts and queries for controlled evaluation, we examine how injecting narrative context into embeddings affects retrieval across abstraction levels.
Concrete queries are retrievable without context, while abstract levels increasingly depend on narrative grounding.
Where context is injected also matters, with image-side enrichment proving particularly effective.
The most abstract level, however, remains challenging even with full context, highlighting context-dependent image retrieval as an important open problem.
Our framework and findings lay groundwork toward retrieval systems that handle the context-dependent meanings images acquire in narrative settings.
\end{abstract}

%%
%% CCS Concepts
%%



\maketitle

%% ============================================
%% 1. Introduction
%% ============================================
\section{Introduction}

\begin{figure}[t]
    \centering
    \begin{minipage}[c]{0.45\linewidth}
        \centering
        \includegraphics[width=\linewidth]{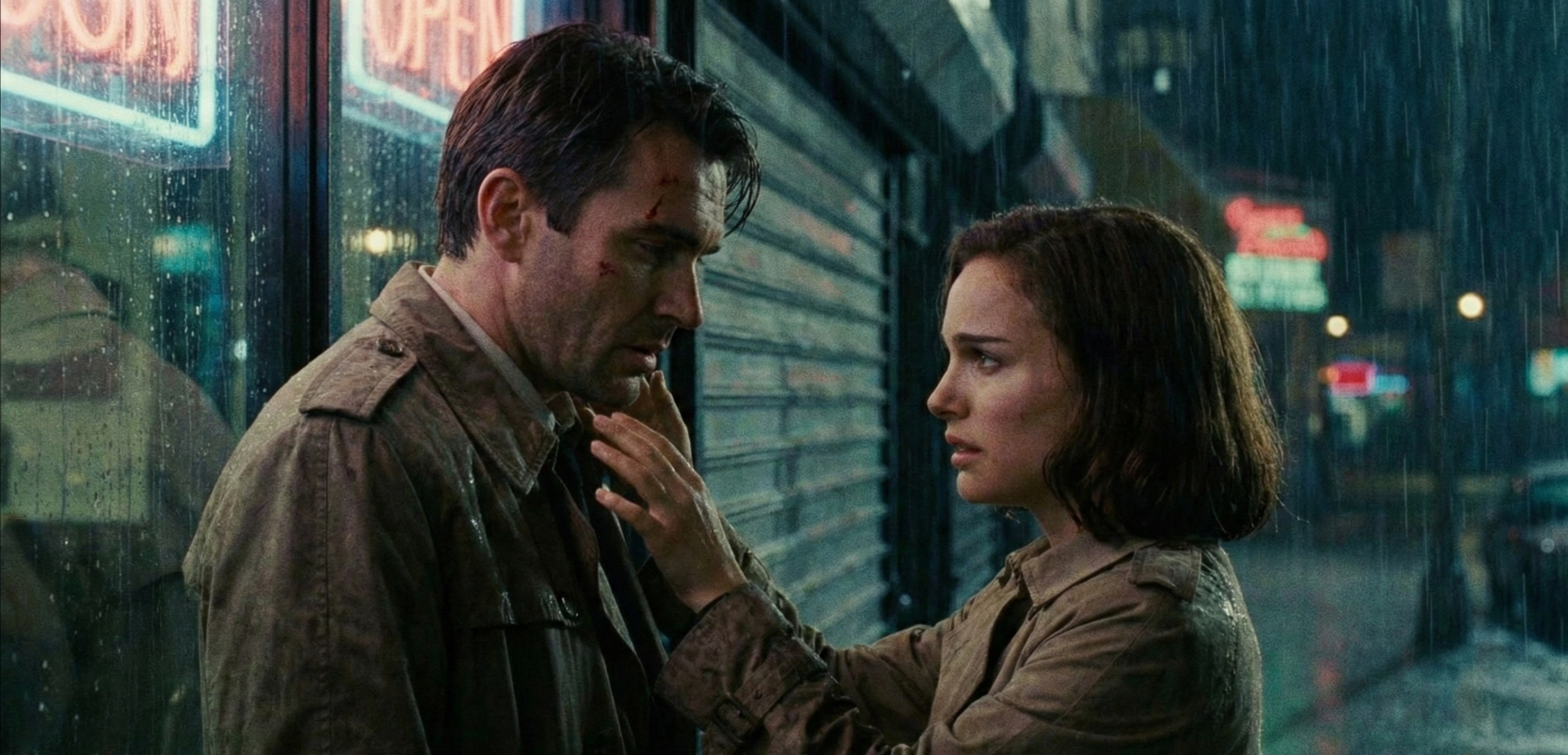}
    \end{minipage}%
    \hfill
    \begin{minipage}[c]{0.50\linewidth}
        \small
        \textcolor{blue!70!black}{$\rightarrow$ \textbf{Reunion}: ``After the Storm''}\\
        {\footnotesize\textcolor{blue!40!black}{A couple reunites after surviving danger.}}\\[0.6em]
        \textcolor{red!70!black}{$\rightarrow$ \textbf{Farewell}: ``The Last Goodbye''}\\
        {\footnotesize\textcolor{red!40!black}{Two people part ways forever.}}
    \end{minipage}

    \medskip
    \footnotesize
    \begin{tabular}{@{}cclll@{}}
        \toprule
        \multicolumn{3}{@{}l}{\textcolor{gray}{Context Dependency}} & \textbf{\textcolor{blue!70!black}{Reunion}} & \textbf{\textcolor{red!70!black}{Farewell}} \\
        \midrule
        \textcolor{gray!50}{$\circ$} & L1 & \textcolor{gray}{Objects, Actions} & \textcolor{gray}{Two people in rain} & \textcolor{gray}{Two people in rain} \\
        \textcolor{gray!70}{$\bullet$} & L2 & \textcolor{gray}{Focus Point} & \textcolor{blue!40!black}{The embrace} & \textcolor{red!40!black}{The distance} \\
        \textcolor{gray}{$\bullet\bullet$} & L3 & \textcolor{gray}{Situation, Intent} & \textcolor{blue!55!black}{Relief after danger} & \textcolor{red!55!black}{Last goodbye} \\
        \textcolor{gray!30!black}{$\bullet\bullet\bullet$} & L4 & \textcolor{gray}{Atmosphere, Effect} & \textcolor{blue!70!black}{Hope, warmth} & \textcolor{red!70!black}{Sorrow, finality} \\
        \bottomrule
    \end{tabular}
    \caption{\small Context shapes interpretation: the same image yields different meanings depending on narrative context. L1 (objects) remains identical; L2--L4 diverge as context dependency increases.}
    \label{fig:context_motivation}
\end{figure}

% P1: 問題提起（文脈の欠如）
Image retrieval has achieved remarkable success for queries about concrete visual elements such as objects, actions, and attributes~\cite{yang2023atomic,wei2024uniir}, driven by models that embed images and text in a shared semantic space~\cite{radford2021clip,li2022blip}.
Yet the full meaning of an image depends not only on its pixels but on the context in which it appears~\cite{huang2016visual}: a scene of two people in the rain can convey hope and warmth in a reunion story, or sorrow and finality in a farewell (Figure~\ref{fig:context_motivation}).
These models, however, map each image to a single fixed point in the shared embedding space, unable to capture how narrative context reshapes its meaning.

% P2: なぜ文脈が重要か（観察と洞察）+ 図

%% --- Example figure: 5 images + L1-L4 × two contexts ---
\begin{figure*}[t]
    \centering
    \setlength{\tabcolsep}{3pt}
    \renewcommand{\arraystretch}{1.1}
    \resizebox{\textwidth}{!}{%
    \small
    \begin{tabular}{@{}cl|p{3.4cm}p{3.4cm}p{3.4cm}p{3.4cm}p{3.4cm}@{}}
    \toprule
    & & \multicolumn{1}{c}{\includegraphics[width=3.2cm]{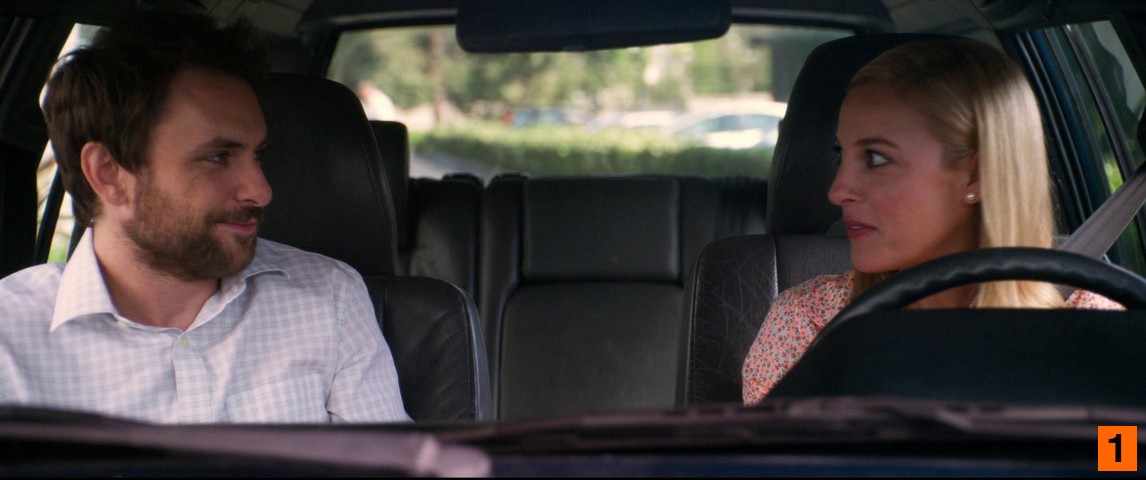}} & \multicolumn{1}{c}{\includegraphics[width=3.2cm]{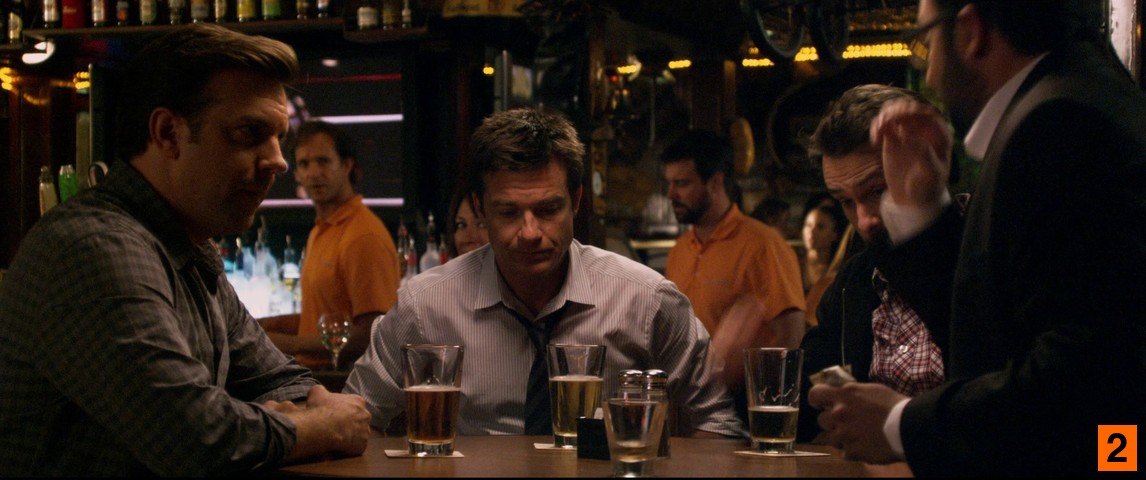}} & \multicolumn{1}{c}{\includegraphics[width=3.2cm]{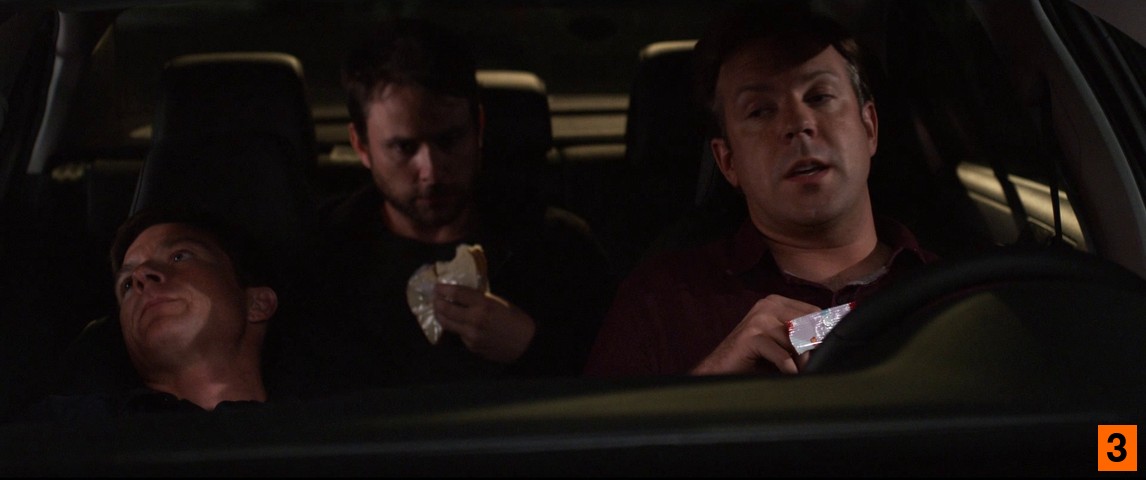}} & \multicolumn{1}{c}{\includegraphics[width=3.2cm]{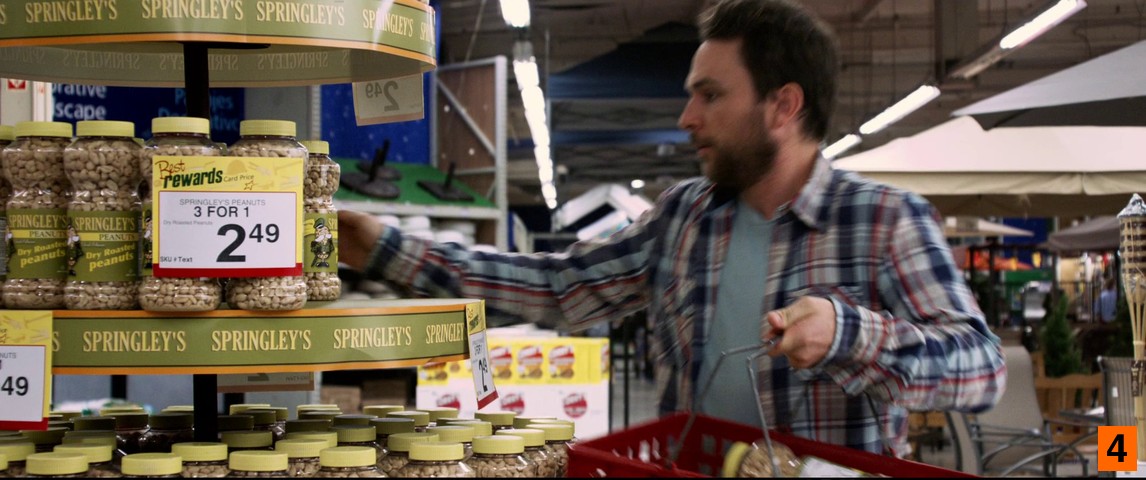}} & \multicolumn{1}{c}{\includegraphics[width=3.2cm]{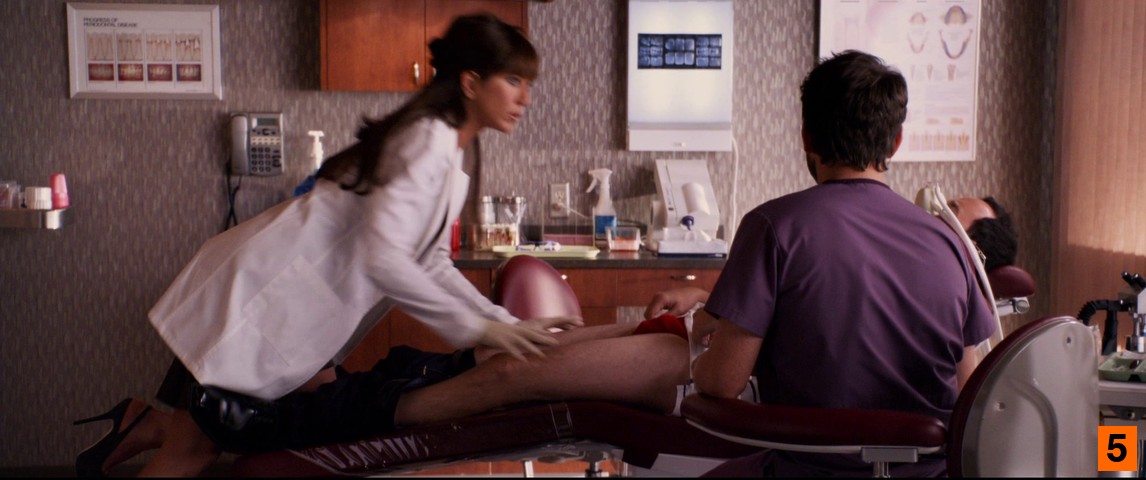}} \\
    \midrule
    \multirow{2}{*}{\textbf{L1}} & $C_1$ & A man and woman in a car; she drives, he looks at her & Three men at a bar with drinks; one looks down, another gestures & Three men in a car at night; driver talks, passenger rests, one eats & A man reaches for jars on a supermarket shelf, holding a basket & A woman in a lab coat examines a man on a medical chair \\
     & $C_2$ & A man and woman in a car; she glances back at him & Three men at a bar with drinks; one looks down, others talk & Three men in a car at night; one drives, one is slumped, one eats & A man in a supermarket aisle reaches for jars, holding a basket & A woman in a lab coat leans over a man on a medical chair \\
    \midrule
    \multirow{2}{*}{\textbf{L2}} & $C_1$ & \textcolor{blue!40!black}{Her warning look; sets up his reckless tendencies} & \textcolor{blue!40!black}{The man's despair spurs friends to hatch a misguided scheme} & \textcolor{blue!40!black}{The driver outlining a flawed ``stealth'' plan} & \textcolor{blue!40!black}{The peanuts he believes are crucial for his absurd mission} & \textcolor{blue!40!black}{The doctor's bewilderment at an embarrassing predicament} \\
     & $C_2$ & \textcolor{red!40!black}{Her feigned concern masks her manipulation of him} & \textcolor{red!40!black}{The man's downward gaze signals deep guilt about the scheme} & \textcolor{red!40!black}{The passenger's sedated, vulnerable state; stripped of agency} & \textcolor{red!40!black}{A frantic search for items to cover up a dark act} & \textcolor{red!40!black}{The woman performing a non-consensual procedure on him} \\
    \midrule
    \multirow{2}{*}{\textbf{L3}} & $C_1$ & \textcolor{blue!55!black}{A girlfriend pleads for sensible behavior before a night out} & \textcolor{blue!55!black}{Friends commiserate and plot a new comedic scheme} & \textcolor{blue!55!black}{Planning a supposedly stealthy retrieval operation} & \textcolor{blue!55!black}{A misguided scavenger hunt for a ``lucky charm''} & \textcolor{blue!55!black}{A medical emergency from a bizarre failed stealth mission} \\
     & $C_2$ & \textcolor{red!55!black}{A deceptive conversation about a dangerous ``investment''} & \textcolor{red!55!black}{A man is coerced further into a dangerous illegal plot} & \textcolor{red!55!black}{Clandestine transport of a drugged man for a procedure} & \textcolor{red!55!black}{A frantic supply run to hide evidence of a crime} & \textcolor{red!55!black}{The mastermind implants a device to ensure the victim's silence} \\
    \midrule
    \multirow{2}{*}{\textbf{L4}} & $C_1$ & \textcolor{blue!70!black}{Light-hearted; anticipation of comedic mishap} & \textcolor{blue!70!black}{Despondent yet hinting at comic absurdity} & \textcolor{blue!70!black}{Farcical anticipation; amusement at misplaced confidence} & \textcolor{blue!70!black}{Comical and absurd; anticipate farcical failure} & \textcolor{blue!70!black}{Absurdly humiliating; shock and laughter} \\
     & $C_2$ & \textcolor{red!70!black}{Subtly tense; foreshadows dark revelation} & \textcolor{red!70!black}{Heavy, conspiratorial; unease about grave consequences} & \textcolor{red!70!black}{Grim and urgent; discomfort and apprehension} & \textcolor{red!70!black}{Anxious, foreboding; dread intensifies} & \textcolor{red!70!black}{Chilling betrayal; dread and vulnerability} \\
    \bottomrule
    \end{tabular}%
    }
    \caption{L1--L4 queries for the same image group under two story contexts. \textbf{$C_1$} (Comedy): ``The Peculiar Peanut Predicament'' --- friends' scheme to retrieve a lucky charm goes hilariously wrong. \textbf{$C_2$} (Thriller): ``The Extraction'' --- a man is trapped in a dangerous scheme orchestrated by his manipulative wife. L1 describes nearly identical visual content; L2--L4 progressively diverge as context dependency increases.}
    \label{fig:example}
\end{figure*}

%% --- Overview figure (separated) ---
\begin{figure*}[t]
    \centering
    \includegraphics[width=0.85\textwidth]{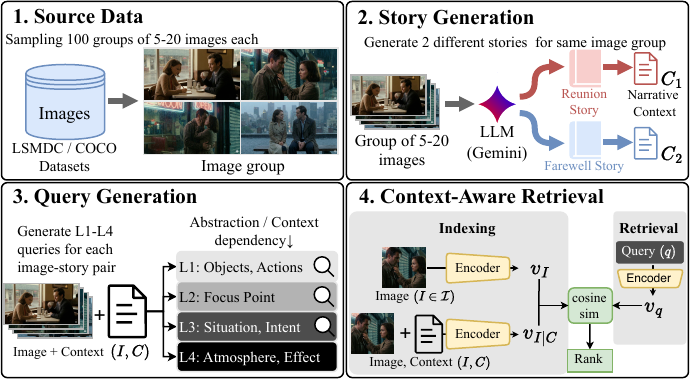}
    \caption{\small Overview of the evaluation pipeline. Steps 1--3 construct narrative contexts $C$ and L1--L4 queries $q$ from image groups; Step 4 evaluates retrieval with and without context injection.}
    \label{fig:overview}
\end{figure*}

Narrative context does not reshape all aspects of meaning equally.
While objects and actions stay identical across stories, what the scene conveys diverges increasingly: focal point (embrace vs.\ distance), situation (relief vs.\ farewell), and atmosphere (hope vs.\ sorrow) each shift further from shared visual content (Figure~\ref{fig:context_motivation}).
This gradient suggests that \textbf{the degree of context dependency correlates with semantic abstraction level}~\cite{zhao2022affective}.
Retrieval systems that ignore context therefore face progressive degradation as queries target more abstract semantics~\cite{pandiani2023seeing}.

% P2.5: 技術的実現可能性
Injecting narrative context into image representations has only recently become feasible, as dual-encoder models such as CLIP~\cite{radford2021clip} encode text and images independently.
Modern VLMs~\cite{chen2024internvl,wang2024qwen2vl} process images alongside textual input via autoregressive decoding; their hidden states can be extracted as context-conditioned embeddings~\cite{behnamghader2024llm2vec}.
E5-V~\cite{jiang2024e5v} and VLM2Vec~\cite{jiang2025vlm2vec} extend this by fine-tuning VLMs for multimodal embedding, integrating image and text semantics in a unified semantic space, enabling context-aware image representations more straightforwardly.

% P3: 本研究のアプローチ
We investigate how well modern multimodal semantic representations capture meaning shifts that intensify with semantic abstraction.
To this end, we introduce the \textbf{L1--L4 framework}, which organizes image semantics into four levels by context dependency: from context-independent (L1) to maximally context-dependent (L4).
Using this framework with synthetic contexts and queries for controlled evaluation, we evaluate multiple model types across context injection strategies, context richness, and abstraction levels.

% P4: Contributions
Our contributions are:
\begin{enumerate}
    \item The L1--L4 framework as an \textbf{analytical lens} for context-dependent image retrieval, revealing a systematic gradient of retrieval difficulty along semantic abstraction levels
    \item Empirical findings on \textbf{context injection strategies}: context dependency forms a gradient along semantic abstraction, image-side context enrichment is particularly effective, and the most abstract level remains challenging even with full narrative context
\end{enumerate}

%% ============================================
%% 2. Approach
%% ============================================
\section{Approach}
\label{sec:approach}

We first formalize the task to clarify what context-aware retrieval requires, then introduce the L1--L4 framework to characterize how context dependency varies with query abstraction.
We then propose a dataset construction method grounded in this formulation and framework, and describe the retrieval methods under evaluation.

\paragraph{Task.}
Given an image collection $\mathcal{I} = \{I_1, I_2, \ldots, I_n\}$, a text query $q$, and a narrative context $C$ that specifies the interpretive setting, the task is to retrieve the most relevant image:
\begin{equation}
I^* = \arg\max_{I \in \mathcal{I}} f(I, q \mid C)
\end{equation}
where $f$ is a scoring function that evaluates how well image $I$ matches query $q$ under context $C$.
Crucially, $f$ must capture meaning that is not inherent in the image but constituted by its context $C$.

\paragraph{L1--L4 Framework.}
To examine how $C$ influences retrieval, we assume $C$ to be a story that situates a group of images in a shared narrative, and operationalize $q$ into four abstraction levels as the L1--L4 framework (Figure~\ref{fig:example}), inspired by settings such as movie scene retrieval from scripts~\cite{tian2025sc2st}.
L1 captures context-independent elements such as objects and actions; L2, focal emphasis determined by narrative attention; L3, situation and intent inferred from the narrative; and L4, atmosphere and emotional effect that arises only from the story context.
This reflects the observation that context dependency appears tied to semantic abstraction: concrete elements are identifiable in isolation, while narrative or emotional interpretations require context to take shape (cf.\@ Figure~\ref{fig:context_motivation}).
Image understanding research corroborates this view: situation recognition extends beyond object identification~\cite{yatskar2016imsitu}, while intent inference~\cite{park2020visualcomet} and affective interpretation~\cite{zhao2022affective} operate at progressively higher and more challenging levels.

% Example + Overview combined figure (placed as figure* after Dataset Construction paragraph)

\paragraph{Dataset Construction.}
We now describe how to construct $(I, C, q)$ triplets, as exemplified in Figure~\ref{fig:example}, at scale.
We use two image sources: \textbf{LSMDC}~\cite{rohrbach2015lsmdc}, which provides movie frames grouped by film, and \textbf{MS-COCO}~\cite{lin2014coco}, which provides independent general images.
From each source, we sample image sequences to form groups (Figure~\ref{fig:overview}, Step~1); for LSMDC, images are drawn from the same movie in temporal order.
For each group, we use Gemini 2.5 Flash to generate multiple story contexts $C$ with distinct genres (Figure~\ref{fig:overview}, Step~2).
Each context contains genre, title, synopses at three granularities (full, three-sentence, and one-sentence), and per-image scene descriptions.
These granularity levels support the context richness analysis in Table~\ref{tab:granularity}.
Given each image-context pair $(I, C)$, the model generates queries $q$ at all four levels, yielding evaluation triplets $(I, C, q)$ (Figure~\ref{fig:overview}, Step~3).
Each query is constructed from two complementary aspects at its abstraction level: L1 combines object description and action; L2, the focal element and its narrative relevance; L3, the situation and character intent; and L4, the atmosphere and its effect on the viewer.
We then evaluate whether this query construction can be traced back via retrieval: given a query $q$ and its associated narrative context $C$, can a model retrieve the corresponding image $I$?
While this approach enables controlled evaluation of context dependency, $C$ and $q$ are synthetically generated; extending evaluation to naturally occurring contexts remains future work.

\paragraph{Retrieval Method.}
We score relevance by cosine similarity between an image embedding $v_I$ and a query embedding $v_q$: $f(I, q \mid C) = \mathrm{sim}(v_I, v_q)$.
The central question is then how to inject the narrative context $C$ into the embedding computation (Figure~\ref{fig:overview}, Step~4).
We compare four strategies by varying which side receives context: \textbf{No-Ctx} ($v_I$, $v_q$), \textbf{Ctx(Q)} ($v_I$, $v_{q|C}$), \textbf{Ctx(I)} ($v_{I|C}$, $v_q$), and \textbf{Ctx(B)} ($v_{I|C}$, $v_{q|C}$), where $v_{I|C}$ and $v_{q|C}$ represent embeddings computed with context $C$ prepended to the image and query inputs, respectively.

We evaluate three model types: (1)~\textbf{CLIP}, a dual-encoder that encodes images and text independently and does not support context injection; (2)~\textbf{VLM}, whose autoregressive hidden states can serve as context-conditioned embeddings~\cite{behnamghader2024llm2vec}; and (3)~\textbf{VLM-Emb}, VLMs fine-tuned directly for multimodal embedding~\cite{jiang2024e5v,jiang2025vlm2vec}.
For VLM and VLM-Emb, we extract the hidden state at the last token position of the final layer and apply $\ell_2$ normalization to obtain the embedding vector, following LLM2Vec~\cite{behnamghader2024llm2vec}.
We inject context by prepending story information (genre, title, synopsis) to the input with a task-specific instruction.\footnote{Ctx(I): ``Based on the story context below, summarize this image in one word. Story Context: Genre: \{genre\} Title: \{title\} Synopsis: \{synopsis\}. Summary above image in one word:''. Ctx(Q): ``Story context: Genre: \{genre\} Title: \{title\} Synopsis: \{synopsis\}. Query: \{query text\}''. Context fields vary with granularity level (Table~\ref{tab:granularity}).}
The ``summarize in one word'' instruction follows the training format of E5-V~\cite{jiang2024e5v}, which uses this constraint to concentrate the model's contextualized representation into the last token position.
Although the instruction targets a single-word output, the full narrative context is processed through the model's attention layers, so the resulting embedding captures context-dependent semantics.
We adopt the same prompt format for VLM to ensure comparability.
Crucially, query generation (Gemini) and retrieval evaluation use separate models, ensuring that we test genuine cross-model retrieval rather than reproduction of generated content.
This separation enables systematic comparison of model types and context injection strategies across L1--L4 abstraction levels, revealing how each factor contributes to context-aware retrieval.

%% ============================================
%% 3. Experiments
%% ============================================
\section{Experiments}

%% --- Combined figure: Query Divergence (left) + Main Results (right) ---
\begin{figure*}[t]
    \centering
    \begin{minipage}[t]{0.25\textwidth}
        \vspace{0pt}
        \centering
        \includegraphics[width=\textwidth]{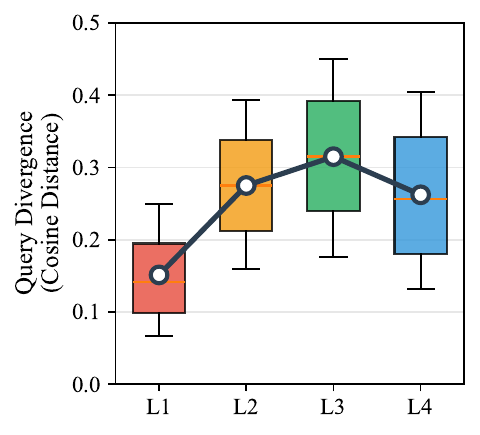}
        \caption{\small Query divergence between two story contexts for the same image (LSMDC+COCO avg.).}
        \label{fig:divergence}
    \end{minipage}%
    \hfill
    \begin{minipage}[t]{0.73\textwidth}
        \vspace{0pt}
        \centering
        \includegraphics[width=\textwidth]{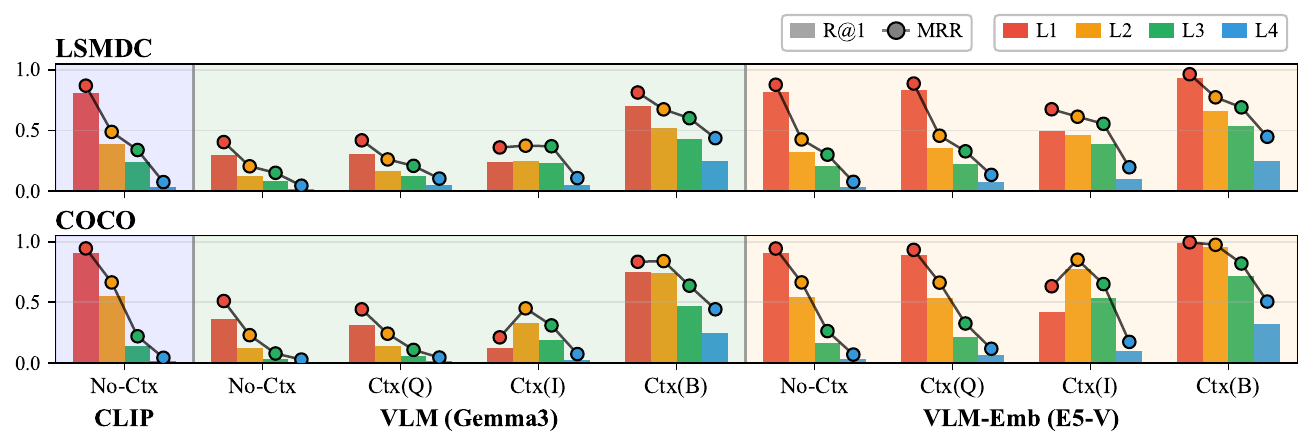}
        \caption{\small Recall@1 by model and context injection pattern for LSMDC (top) and COCO (bottom).}
        \label{fig:main}
    \end{minipage}
\end{figure*}

\subsection{Experimental Setup}\label{sec:setup}

For each dataset (LSMDC, MS-COCO), we construct 100 image groups, each paired with two distinct stories.
Group size varies across 5, 10, and 20 images, yielding retrieval pools of 1{,}000, 2{,}000, and 4{,}000 images respectively.
We report results with 10 images per group, as trends are consistent across sizes.
Queries are ranked against all images in the dataset.
We report Recall@1 as the primary metric, with Mean Reciprocal Rank (MRR) for additional analysis.

For \textbf{CLIP}, we use EVA-CLIP-18B~\cite{sun2023evaclip}; for \textbf{VLM}, Gemma3-4B-PT~\cite{gemmateam2025gemma3}; for \textbf{VLM-Emb}, E5-V~\cite{jiang2024e5v}.
These are the best-performing models identified through preliminary experiments; trends across abstraction levels were consistent across models within each type.

Figure~\ref{fig:divergence} illustrates query divergence, the cosine distance between query embeddings generated under two different story contexts for the same image.
Divergence rises sharply from L1 to L2--L4 ($p < .001$, Mann--Whitney), though L4 falls below L3, likely because highly abstract descriptions converge on a limited emotional vocabulary.

\subsection{Results}

%% --- Main Results: Model × Context Injection Pattern ---
\paragraph{Main Results.}
Figure~\ref{fig:main} shows Recall@1 across abstraction levels for each model and context injection pattern.
Without context injection (No-Ctx), performance declines sharply from L1 to L4 across all models, consistent with the expectation that abstract queries require grounding beyond visual features alone.
This monotonic decline reveals a gradient of context dependency aligned with semantic abstraction.

Context injection substantially improves L2--L4, but the injection side matters.
Ctx(Q) yields limited improvement, as query-side injection alone cannot enrich image representations.
Ctx(I) improves L2--L4 but degrades L1, because injecting context shifts image embeddings away from literal visual features.
Ctx(B) achieves the best overall performance by enriching both sides, preserving L1 while substantially improving abstract levels.
The asymmetry between Ctx(Q) and Ctx(I) indicates that context must be injected where the semantic gap is greatest---the image side.
Yet even with Ctx(B), L4 retrieval remains largely unsolved, confirming that atmosphere and symbolic meaning pose a genuinely open challenge.

VLM-Emb (E5-V) consistently outperforms VLM (Gemma3), indicating that embedding-oriented fine-tuning is critical for context-aware retrieval---raw autoregressive hidden states do not yet produce retrieval-quality representations for abstract semantics.
Subsequent analyses adopt E5-V with Ctx(B) as the strongest configuration and report Recall@1 on LSMDC+COCO averages, as MRR shows consistent trends.
We next examine three aspects: how much context is needed, whether explicit level specification helps, and how ranking priorities shift across abstraction levels.

% ---- Granularity Analysis ----
\paragraph{Context Granularity.}
We vary the granularity of injected context, from genre metadata alone to full plot synopses, to examine how much contextual detail is needed for effective retrieval at each abstraction level.
Table~\ref{tab:granularity} reports the results across six granularity levels (g1--g6).
L1 shows minimal dependency on context granularity (+.03 to +.08 from g1 to g6), as visual features suffice for concrete queries.
L2--L3 exhibit the largest gains: Ctx(I) improves by +.34 at L2 and +.33 at L3; Ctx(B) shows +.31 and +.39 respectively.
L4 benefits moderately, with Ctx(B) gaining +.22 while Ctx(I) shows limited improvement (+.07).
Both injection patterns benefit from richer context overall, though their improvement trajectories differ: Ctx(B) achieves large initial gains at g2 (genre+title) and continues to improve more gradually, while Ctx(I) improves steadily up to g6 (full synopsis).
When both sides share context, even minimal metadata serves as a semantic anchor that accelerates early gains; when only the image side carries context, richer narrative detail is needed to compensate for the absence of query-side alignment.

\begin{table}[t]
    \caption{\small Recall@1 by context granularity (LSMDC+COCO average). g=genre, t=title, s1/s3/sf=1-sentence/3-sentence/full synopsis ($\dagger$: main result setting).}
    \label{tab:granularity}
    \small
    % tab:granularity_avg - Context Granularity (LSMDC+COCO Average)
% Images: 10, Pool: all, Model: E5V
% Source: vlm-emb_e5v_*_all_*img_ctx[IB]_gX.json
% $\dagger$: Ctx setting used in main result
% Auto-generated by generate_paper_tables.py
\begin{tabular}{@{}l|cccc|cccc@{}}
\toprule
 & \multicolumn{4}{c|}{Ctx(I)} & \multicolumn{4}{c}{Ctx(B)} \\
Granularity & L1 & L2 & L3 & L4 & L1 & L2 & L3 & L4 \\
\midrule
g1 (genre) & .43 & .28 & .13 & .03 & .88 & .50 & .24 & .06 \\
g2 (title) & .40 & .29 & .16 & .03 & .93 & .67 & .41 & .23 \\
g3 (g+t) & .41 & .31 & .17 & .03 & .96 & .72 & .44 & .24 \\
g4 (+s1) & .39 & .40 & .26 & .05 & .97 & .78 & .52 & .23 \\
g5 (+s3) & .43 & .54 & .39 & .08 & .97 & .79 & .56 & .23 \\
g6 (+sf) $\dagger$ & .46 & .62 & .46 & .10 & .96 & .81 & .63 & .28 \\
\bottomrule
\end{tabular}
    \end{table}

% ---- Level-aware Analysis ----
\paragraph{Level-Aware Prompting.}
We test whether explicitly specifying the target abstraction level in prompts improves retrieval, by prepending level-specific focus instructions (e.g., ``Focus on objects and actions'' for L1) to guide the model toward the appropriate degree of abstraction.
Table~\ref{tab:levelaware} shows the results.
However, level-aware prompting yields only marginal gains (+1--3\%), suggesting that VLM embeddings already capture abstraction degree implicitly, without explicit level specification.
If so, models should exhibit different discrimination behaviors across abstraction levels even without level labels---we examine this next.

\begin{table}[t]
    \caption{\small Level-aware (LA) prompting effect (LSMDC+COCO average).}
    \label{tab:levelaware}
    \small
    % tab:levelaware_wide - Level-Aware Effect (Wide Format)
% Images: 10, Pool: all, Model: E5V, LSMDC+COCO Average
% Auto-generated by generate_paper_tables.py
\begin{tabular}{@{}l|cccc|cccc|cccc@{}}
\toprule
LA & \multicolumn{4}{c|}{No-Ctx} & \multicolumn{4}{c|}{Ctx(I)} & \multicolumn{4}{c}{Ctx(B)} \\
 & L1 & L2 & L3 & L4 & L1 & L2 & L3 & L4 & L1 & L2 & L3 & L4 \\
\midrule
N & .86 & .43 & .19 & .03 & .46 & .62 & .46 & .10 & .96 & .81 & .63 & .28 \\
Y & -- & -- & -- & -- & .47 & .63 & .47 & .11 & .97 & .81 & .62 & .31 \\
\bottomrule
\end{tabular}
    \end{table}

% ---- Discrimination Patterns ----
\paragraph{Discrimination Patterns.}
\begin{figure}[t]
    \centering
    \includegraphics[width=\columnwidth]{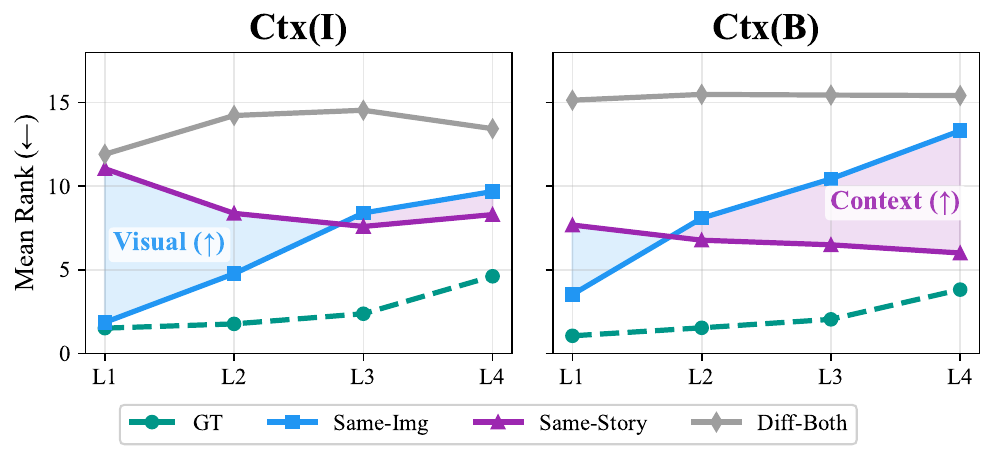}
    \caption{\small Mean rank (lower = ranked higher) by image category within each group. Four categories: GT (correct image), Same-Img (same image, different story), Same-Story (different image, same story), Diff-Both (different image and story).}
    \label{fig:discrimination}
    \end{figure}

To understand how context injection reshapes rankings beyond Recall@1, we analyze within-group ranking (10 images per group).
Because each group is paired with two stories (Section~\ref{sec:setup}), candidates fall into four categories: GT (correct image), Same-Img (same image, different story), Same-Story (different image, same story), and Diff-Both (different image and story).
Their mean ranks reveal whether the model prioritizes visual or narrative similarity.

Figure~\ref{fig:discrimination} plots mean rank for these categories across L1--L4, with separate panels for Ctx(I) and Ctx(B).
Across all levels, GT consistently ranks highest and Diff-Both lowest, confirming that the model correctly identifies the target and demotes unrelated images.
The key finding is the crossover between Same-Img and Same-Story: at L1, Same-Img ranks close to GT (visual similarity dominates), but as abstraction increases, Same-Story rises while Same-Img falls, crossing over at L3--L4 as the discrimination criterion shifts from visual appearance to narrative context.
At L4, however, Same-Story approaches GT in rank, explaining why L4 Recall@1 remains low: the model captures narrative similarity but struggles to discriminate individual images within the same story at the level of atmosphere and symbolic meaning.

%% ============================================
%% 4. Related Work
%% ============================================
\section{Related Work}

Representing images and text in a shared semantic space has been central to image understanding and retrieval.
Contrastive dual-encoders such as CLIP~\cite{radford2021clip} mapped each image to a single fixed embedding, limiting compositional reasoning and context-dependent interpretation~\cite{yuksekgonul2023aro,thrush2022winoground}.
Vision-language models (VLMs)~\cite{chen2024internvl,wang2024qwen2vl} moved beyond this limitation by processing images alongside text via autoregressive decoding, enabling context-conditioned representations~\cite{behnamghader2024llm2vec}.
E5-V~\cite{jiang2024e5v} and VLM2Vec~\cite{jiang2025vlm2vec} further fine-tuned these VLMs for multimodal embedding, producing retrieval-oriented representations that integrate image and text in a unified space.

Incorporating auxiliary textual information into image retrieval queries is a closely related research direction, formulated as Composed Image Retrieval (CIR).
Methods such as Pic2Word, SEARLE, and MagicLens~\cite{saito2023pic2word,baldrati2023searle,zhang2024magiclens} combined a reference image with textual modifications to retrieve visually altered targets.
However, CIR focuses on explicit visual changes such as attributes or objects rather than abstract, context-dependent semantics such as mood or narrative intent, and operates via query-side augmentation, whereas our approach injects context on the image side to enrich representations themselves.
Context-aware retrieval benchmarks such as ImageCoDe~\cite{krojer2021imagecode} and CoVR~\cite{ventura2024covr} acknowledged the role of context, but did not systematically inject rich narrative context into image embeddings or evaluate across different levels of semantic abstraction.

Beyond retrieval, higher-level image understanding that involves inferring intents, causes, and emotional effects is central to what our L3--L4 levels aim to capture.
Visual commonsense reasoning tasks such as VisualCOMET~\cite{park2020visualcomet} and VCR~\cite{zellers2019vcr} addressed such understanding, yet they targeted question-answering rather than retrieval; similarly, visual sentiment analysis captured abstract mood but assumed a fixed interpretation per image.
Our work is, to our knowledge, the first to systematically evaluate how context granularity and injection patterns affect retrieval performance across semantic abstraction levels, using the L1--L4 framework to reveal the gradient of context dependency.

%% ============================================
%% 5. Conclusion
%% ============================================
\section{Conclusion}

We studied context-dependent image retrieval across semantic abstraction levels using the L1--L4 framework.
Our key findings are:
\begin{itemize}
    \item Context dependency aligns with semantic abstraction: concrete semantics (L1) are retrievable without context, while abstract levels (L2--L4) increasingly depend on narrative grounding that visual features alone cannot provide.
    \item Image-side context injection is particularly effective, and injecting context on both sides best balances concrete and abstract retrieval.
    \item Atmosphere and symbolic meaning (L4) remain largely unretrievable even with full context, pointing to fundamental limitations in current VLM representations.
\end{itemize}
These results suggest that context-aware retrieval can enable applications beyond visual similarity, such as retrieving movie scenes by narrative function or searching personal photo archives by what moments meant within a trip or life event.
In such settings, image-side context injection (Ctx(I)) offers a practical deployment path, as narratives can be associated with images at indexing time without modifying query processing.
However, our evaluation relies on synthetically generated contexts and queries; validating these findings with naturally occurring narratives, such as real film scripts or user-authored captions, and closing the L4 gap remain important next steps.

%% ============================================
%% References
%% ============================================
\bibliographystyle{ACM-Reference-Format}
\bibliography{references}

\end{document}